\documentclass[reprint,amsmath,aps]{revtex4-2}
\usepackage{aasmacros} 
\usepackage{changes}
\usepackage{graphicx}
\usepackage{dcolumn}
\usepackage{bm}
\usepackage{hyperref} 

\usepackage{ulem}

\newcommand{\hMpc}{{\ifmmode{\,h^{-1}{\rm Mpc}}\else{$h^{-1}$Mpc}\fi}}
\newcommand{\hMpcc}{{\ifmmode{\,h\,{\rm Mpc}^{-1}}\else{$h$\,Mpc$^{-1}$}\fi}}
\newcommand{\Mpcc}{{\ifmmode{\,{\rm Mpc}^{-1}}\else{\,Mpc$^{-1}$}\fi}}

\newcommand{\hkpc}{{\ifmmode{\,h^{-1}{\rm kpc}}\else{$h^{-1}$kpc}\fi}}
\newcommand{\hMsun}{{\ifmmode{\,h^{-1}{\rm {M_{\odot}}}}\else{$h^{-1}{\rm{M_{\odot}}}$}\fi}}
\newcommand{\Msun}{\,\rm {M_{\odot}}}
\newcommand{\Mstar}{{\ifmmode{\,M_{*}}\else{$M_{*}$}\fi}}
\newcommand{\Mhalo}{{\ifmmode{\,M_{\rm halo}}\else{$M_{\rm halo}$}\fi}}
\newcommand{\ltsima}{$\; \buildrel < \over \sim \;$}
\newcommand{\gtsima}{$\; \buildrel > \over \sim \;$}
\newcommand{\lsim}{\lower.5ex\hbox{\ltsima}}
\newcommand{\gsim}{\lower.5ex\hbox{\gtsima}}

\newcommand{\wen}[1]{\textcolor{blue}{#1}}

\begin{document}
	
\preprint{APS/123-QED}

\title{Constraining neutrino mass with the CSST galaxy clusters}

\author{Mingjing Chen}
	\email{mingjing@mail.ustc.edu.cn}
	\affiliation{CAS Key Laboratory for Research in Galaxies and Cosmology, Department of Astronomy, University of Science and Technology of China, Hefei, Anhui, 230026, People’s Republic of China}
	\affiliation{School of Astronomy and Space Science, University of Science and Technology of China, Hefei, Anhui, 230026, People’s Republic of China}
\author{Yufei Zhang}
	\email{zyfeee@mail.ustc.edu.cn}
    	\affiliation{College of Mathematics and Physics, Leshan Normal University, 614000, China}

     \affiliation{CAS Key Laboratory for Research in Galaxies and Cosmology, Department of Astronomy, University of Science and Technology of China, Hefei, Anhui, 230026, People’s Republic of China}
    	\affiliation{School of Astronomy and Space Science, University of Science and Technology of China, Hefei, Anhui, 230026, People’s Republic of China}
\author{Zhonglue Wen}
	\email{zhonglue@nao.cas.cn}
	\affiliation{CAS Key Laboratory of FAST, NAOC, Chinese Academy of Sciences, Beijing 100101, People’s Republic of China}
	\affiliation{National Astronomical Observatories, Chinese Academy of Sciences, 20A Datun Road, Chaoyang District, Beijing 100101, People’s Republic of China}
\author{Weiguang Cui}
    \thanks{Talento-CM fellow}
    \email{weiguang.cui@uam.es}
    \affiliation{Departamento de F\'{i}sica Te\'{o}rica, Universidad Aut\'{o}noma de Madrid, M\'{o}dulo 15, E-28049 Madrid, Spain}
    \affiliation{Centro de Investigaci\'{o}n Avanzada en F\'isica Fundamental (CIAFF), Facultad de Ciencias, Universidad Aut\'{o}noma de Madrid, 28049 Madrid, Spain}
    \affiliation{Institute for Astronomy, University of Edinburgh, Royal Observatory, Edinburgh EH9 3HJ, UK}
\author{Wenjuan Fang}
	\email{wjfang@ustc.edu.cn}
	\affiliation{CAS Key Laboratory for Research in Galaxies and Cosmology, Department of Astronomy, University of Science and Technology of China, Hefei, Anhui, 230026, People’s Republic of China}
	\affiliation{School of Astronomy and Space Science, University of Science and Technology of China, Hefei, Anhui, 230026, People’s Republic of China}

\date{\today}

\begin{abstract}
With the advent of next-generation surveys, constraints on cosmological parameters are anticipated to become more stringent, particularly for the total neutrino mass. This study forecasts such constraints utilizing galaxy clusters from the Chinese Space Station Telescope (CSST). 
Employing Fisher matrix techniques, we derive the constraint $\sigma(M_\nu)$ from cluster number counts,  cluster power spectrum, and their combination. The investigation covers both the standard cosmological model with massive neutrinos $\nu\Lambda$CDM and the inclusion of dynamical dark energy in the $\nu w_0 w_a$CDM model, revealing a minor impact of dark energy on neutrino mass constraints.
We examine the largest source of systematic arising from the mass-observable relation uncertainties and find that, with perfect knowledge of the scaling relation parameters, CSST clusters have the potential to enhance precision, tightening constraints to $\sim0.03$ eV. We also study the effects of the maximum redshift $z_{max}$ and other uncertainties, including those in redshift, halo mass function, and bias. Furthermore, we emphasize the significance of accounting for the growth-induced scale-dependent bias (GISDB) effect, which we find can tighten the final constraint by a factor of $1.2$ - $2.2$.

\end{abstract}

\maketitle

\section{introduction}\label{sec01}

In the standard model of particle physics, neutrinos are expected to be massless. However, a significant breakthrough in understanding neutrinos occurred in 1998 when the Super-Kamiokande atmospheric neutrino experiment in Japan confirmed neutrino oscillations \cite{Fukuda1998, Sudbury2016}, that is, neutrinos can change from one flavor to another during their propagation. This discovery provides evidence for non-zero neutrino masses, although small. 

Various oscillation experiments, including those involving solar, atmospheric, accelerator and reactor neutrinos, have been conducted. They can determine the differences in squared masses between the three different mass eigenstates of neutrinos. However, they cannot determine the absolute value of neutrino masses. As a result, neutrino mass hierarchy has two possible scenarios: the normal hierarchy (NH), where the two lighter neutrinos are closer in mass, or the inverted hierarchy (IH), where the two heavier neutrinos are. The current neutrino oscillation experiments \cite{Aiello2022} measured:
\begin{gather*}
	\Delta m_{21}^2 \equiv m_2^2-m_1^2=7.5\times 10^{-5}eV^2  \\
	|\Delta m_{31}^2| \equiv |m_3^2-m_1^2| =2.5\times 10^{-3}eV^2,
\end{gather*}
where the subscript \{1,2,3\} represents different mass eigenstate. So, the oscillation experiments constrain the total neutrino mass $ M_{\nu} \equiv \sum_{\nu=1,2,3}m_{\nu}\gsim0.06$ eV for NH ($m_3>m_2>m_1$), and $ M_{\nu}\gsim0.10 $ eV for IH ($m_2>m_1>m_3$).

How to measure the absolute neutrino mass? There are three different approaches \cite{Drexlin2013}\cite{Gerbino2017}:  (i) neutrinoless double $\beta$-decay $(0\nu \beta \beta)$ searches: the KamLAND-Zen experiment obtained the lightest neutrino mass $ m_{\text{lightest}}<0.18-0.48$ eV at the 90\% confidence level (C.L.) \cite{Gando2016}; (ii) kinematic measurements: the KATRIN tritium decay experiment obtained $ M_{\nu}<1.1$ eV at 90\% C.L. \cite{Aker2021}. (iii) cosmology (see \cite{Lesgourgues2006, Lesgourgues2013} for reviews).

Fixing the matter density today, if neutrinos have mass, the non-relativistic component will decrease in the early Universe because neutrinos exhibit relativistic behavior at that time. This will lead to a delay in the time of matter-radiation equality. Consequently, this delay suppresses the growth of perturbations since the majority of growth takes place during the matter-dominated era.
Additionally, due to their large thermal velocities, neutrinos tend to freely stream out of overdense regions, which is known as the free streaming effect. This effect tends to smear out fluctuations on scales smaller than the free-streaming scale $k_{\text{fs}}$ \cite{Lesgourgues2006}: 
\begin{align}\label{eq:kfs}
	k_{\text{fs}} = 0.82\frac{\sqrt{\Omega_{\Lambda}+\Omega_m(1+z)^3}}{(1 + z)^2} \left( \frac{m_{\nu}}{1eV} \right) \hMpcc,
\end{align}
where $\Omega_{\Lambda}$ (resp. $\Omega_m$) is the cosmological constant (resp. matter) density fraction evaluated today. 

Considering the impacts of neutrinos on the expansion history of the Universe and the growth of cosmic structures, various probes are developed to constrain the total neutrino mass, including the cosmic microwave background (CMB) anisotropies, CMB lensing, galaxy clustering, galaxy lensing, Lyman-$\alpha$ forest, 21 cm intensity mapping (see \cite{Abazajian2011, Abazajian2015} for reviews).
The current tightest constraint comes from combining baryon acoustic oscillation (BAO) data from the Dark Energy Spectroscopic Instrument (DESI), with CMB anisotropies and CMB lensing data from both Planck and Atacama Cosmology Telescope (ACT), resulting in an upper limit of $M_{\nu}<0.072(0.113) $ eV at a 95\% C.L. for a $M_{\nu}>0(0.059) $ eV prior \cite{DESI2024}.
Many studies have been conducted to forecast constraints that can be achieved by next-generation surveys such as the Vera Rubin Observatory Legacy Survey of Space and Time (LSST) \cite{LSST2009},  Nancy Grace Roman Space Telescope \cite{Roman2015} and Square Kilometre Array (SKA)\cite{Maartens2015}.
For instance, Yu et al. \cite{Yu2023} find that a combination of LSST galaxies and CMB-S4 lensing can achieve constraints of $0.025 $ eV on $ M_{\nu}$.
In this study, we focus on the probe of neutrino mass by using galaxy clusters \citep[see][for reviews]{Kravtsov2012, Allen2011} and provide a forecast for the Chinese Space Station Telescope (CSST), a next-generation space-based galaxy survey\cite{Zhan2011, Gong2019}.

As the largest gravitationally bound structures in the Universe, galaxy clusters, although less abundant than galaxies, have more accurate theoretical descriptions and are sensitive to neutrino mass.
Specifically, the two most studied observables of clusters are: cluster number counts $N(z)$ and cluster power spectrum $P_c(k)$. These two are complementary\cite{Wang2005}: massive neutrinos not only suppress the overall amplitude of matter fluctuations, leading to a decrease in $N(z)$, but also affect the fluctuations on scales smaller than the free-streaming scales, consequently influencing the shape of $P_c(k)$.

CSST is a 2-meter space telescope operating in the same orbit as the China Manned Space Station. It will perform simultaneously a multi-band imaging and a slitless spectroscopic survey with an area of 17,500 deg$^2$ in about ten years. 
Compared to other large-field surveys, CSST excels in image quality, number of filter bands, and its unique coverage in the near ultraviolet etc\cite{Zhan2021}. Hence, CSST is expected to be highly competitive. It will also be complementary to other large projects of its time, such as Euclid \cite{Laureijs2011} and the Roman Space Telescope \cite{Roman2015}. As a result, we expect CSST can provide more accurate constraints on the total neutrino mass as well.

In this work, we aim at forecasting constraints on the total neutrino mass $M_{\nu}$ by using CSST galaxy clusters, specifically the two cluster probes of cluster number counts and power spectrum. The rest of this paper is organized as follows: in section \hyperref[sec:theory]{II}, we introduce our methodology, which is based on the Fisher matrix technique, with careful incorporation of effects from massive neutrinos. Section \hyperref[sec:results]{III} presents our main results for different models and assumptions. In Section \hyperref[sec:discussions]{IV}, we discuss various effects that could impact our findings. Finally, we summarize our results and conclude in Section \hyperref[sec:conclusions]{V}.

Throughout this paper, natural logarithm is denoted as $\ln$, while $\log$ stands for decimal logarithm. The halo mass is defined as $M\equiv M_{200m}$, i.e. cluster mass enclosed within a radius with an interior overdensity of \textit{200} times the \textit{mean} matter density, unless otherwise specified.

\section{theoretical calculations}\label{sec:theory}
In this section, we present our theoretical calculations for cluster number counts and cluster power spectrum, as well as their Fisher matrices. Particularly, we incorporate the effects of massive neutrinos on both observables.

\subsection{Cluster number counts}
\label{sec:N}

Galaxy clusters can be taken as dark matter halos, with their comoving number density $n$ described by the halo mass function (hereafter, HMF), denoted as $d n/d \ln M (M, z)$, as a function of cluster mass $\ln M$ and redshift $z$:
\begin{equation}\label{eq:hmf}
	\frac{d n}{d \ln M} (M, z) = \frac{\rho_m}{M} \cdot f (\sigma) \left| \frac{d \ln \sigma}{d \ln M} \right|.
\end{equation}
Here, $f(\sigma)$ is a function calibrated against N-body simulations, which is related to the nonlinear collapse \cite{PS1974, ST1999, SMT2001,Tinker2008,Tinker2010}. $\rho_m$ represents the matter density at the present day. 
The quantity $\sigma(R,z)$ is the amplitude of the linear matter fluctuations at redshift $z$, smoothed by a top hat window function with scale $R$, and can be expressed as:
\begin{equation}\label{eq:sigmaR}
	\sigma^2 (R, z) = \frac{1}{2 \pi^2} \int_0^{\infty} k^2 P_m (k, z) W^2 (k R) dk,
\end{equation}
where, $W (k R)$ represents the Fourier transform of the top-hat window function with a radius $R$, given by 
\begin{equation}\label{eq:WkR}
	W (k R) = \frac{ 3 [\sin (k R) - k R \cos (k R)]}{(k R)^3}.
\end{equation}
The relationship between the halo mass $M$ and the scale $R$ in the top-hat window function is given by:
\begin{equation}\label{eq:MR}
	M = \frac{4\pi}{3} R^3 \rho_m.
\end{equation}
The linear matter power spectrum $P_m$ in Eq.\eqref{eq:sigmaR} is obtained from CLASS (Cosmic Linear Anisotropy Solving System) \cite{Blas2011}.

%
%

By integrating Eq.\eqref{eq:hmf}, we can determine the abundance of clusters within a specific mass and redshift range $(\Delta \ln M, \Delta z)$:
\begin{equation}\label{eq:N1}
	N(\Delta \ln M, \Delta z) = \Delta \Omega  \int_{ \Delta z} dz \frac{dV (z)}{dzd \Omega} \int_{\Delta \ln M} \frac{dn (M, z)}{d \ln M} d\ln M,
\end{equation}
where, $\Delta \Omega$ represents the solid angle covered by the survey, and $dV(z)/dzd\Omega$ denotes the comoving volume per unit redshift and solid angle intervals.

%
%
Considering an actual survey, there is always an error between the true and observed redshifts $[z,z^{ob}]$, as well as for cluster mass $[M,M^{ob}]$. For the observed redshift, we employ a Gaussian distribution for $z^{ob}$, $P(z^{ob} \mid z)$, with an expected value of $z$ and a variance of $\sigma^2_{z^{ob} \mid z}$:
\begin{equation}\label{eq:Pz}
	P (  z^{ob} \mid z) 
        = \frac{1}{\sqrt{2 \pi} 
        	\sigma_{ z^{ob} \mid  z}} 
        \exp \left[ - \frac{( z^{ob}- z)^2}{2 \sigma^2_{ z^{ob} \mid  z}}
        \right],
\end{equation}
\begin{equation}\label{eq:sigmaz}
	\sigma_{ z^{ob} \mid  z}( z)=\sigma_z(1+ z), 
\end{equation}
where the dispersion of observed redshift $\sigma_{z^{ob} \mid z}(z)$ generally increases with redshift with a factor of $1+z$ \cite{Cao2018}, reflecting poorer redshift estimates for faint galaxies at higher redshifts.

%
%
For the observed cluster mass, following \cite{Zhang2023} \cite{Sartoris2016}, we assume that the probability distribution function of the observed mass $M^{ob}$ for clusters with a fixed true mass $M$ and true redshift $z$ is given by a log-normal distribution with an expected value $\left<\ln M^{ob}\right>(\ln M ,z)$, and a variance $\sigma^2_{\ln M^{ob} \mid \ln M,z}$:
\begin{align}\label{eq:Pm}
		P (\ln M^{ob} \mid \ln M,z)\ 
		&  = \frac{1}{\sqrt{2 \pi}  \sigma_{\ln M^{ob} | \ln M,z}}  \notag
		\\
		& \times \exp \left[   -    \frac{  x^2 (\ln M^{ob}, \ln M,z)  }{  2 \sigma_{\ln M^{ob} \mid \ln M,z}^2  }      \right],
\end{align}
where $x (\ln M^{ob}, \ln M,z) \equiv \ln M^{ob} - \left<\ln M^{ob}\right>(\ln M ,z)$ represents the deviation of the observed mass from its mean value. The mean value $\left<\ln M^{ob}\right>(\ln M ,z)$ is given by:
\begin{equation}\label{eq:Mmean}
 \left<\ln M^{ob}\right>(\ln M ,z)=\ln M +\ln M_\text{bias},
\end{equation}
where $\ln M_\text{bias}$ is the bias in the logarithm of mass, parameterized as \cite{Sartoris2016}:
\begin{equation}\label{eq:Mbias}
 \ln M_\text{bias} = B_{M,0}+ \alpha \ln(1+z).
\end{equation}
The parameters $B_{M,0}$ and $\alpha$ here will be simultaneously constrained with cosmological parameters, often referred to as the self-calibration scheme.

As for the scatter in the observed mass $\sigma_{\ln M^{ob}}$ in Eq.\eqref{eq:Pm}, we estimate it from the scatter in the observed cluster richness $\sigma_{\ln \lambda^{ob}}$. Richness, denoted by $\lambda$, is a direct observable in optical surveys like CSST, and its relation with mass, i.e. the mass-richness relation, has been studied in many works (e.g. \cite{Simet2017, Murata2019, Capasso2019, Bleem2020}).
Costanzi et al. \cite{Costanzi2021} calibrates this relation with the Dark Energy Survey (DES) year 1 cluster abundance and South Pole Telescope (SPT) multiwavelength data.
They employ a log-normal function for the intrinsic (true) richness distribution, incorporating a constant term and a Poisson-like term to model the variance in the intrinsic (true) richness $\sigma_{\ln \lambda}^2=D_{\lambda}^2+(\left< \lambda(M) \right>-1)/\left< \lambda(M) \right>^2$. 
Additionally, they consider the projection effects -i.e., the influence of correlated and uncorrelated line-of-sight structures on photometric cluster richness estimates (or other observable mass proxy)-by using the same methods as \cite{Costanzi2019_projection} to characterize its impact on the observed richness, i.e., $\lambda \rightarrow \lambda^{ob}$.

However, a fitting result for $\sigma_{\ln \lambda^{ob}}$ is not available; only the mean $\left<\ln \lambda^{ob} \right>(M,z)$ is given:
\begin{align}\label{eq:MRR}
	\left<\ln \lambda^{ob} \right>(M,z)
	&=\ln A_{\lambda} + B_{\lambda} \ln(\frac{M}{3\times 10^{14}M_\odot h^{-1}})   \nonumber
	\\
	& + C_{\lambda} \ln(\frac{1+z}{1+0.45}).
\end{align}
Without the projection effects, we would estimate $\sigma_{\ln \lambda^{ob}}^2$ also as the sum of a constant term, which we denote as $D_{\lambda}^2$ as well, and a Poisson term. That is,
\begin{equation}\label{eq:sigmalambda}
	\sigma_{\ln \lambda^{ob}}^2=D_{\lambda}^2+\frac{1}{\exp   \left<\ln \lambda^{ob}(M) \right>},
\end{equation}
which after scaled by the square of the scaling-relation slope $B_{\lambda}^2$, will lead to a scatter in the logarithm of mass. Finally, we model $\sigma_{\ln M^{ob}}^2$ as the direct sum of this scaled scatter and an extra term accounting for the projection effects:
\begin{equation}\label{eq:sigmaMob}
	\sigma_{\ln M^{ob}}^2= \frac{\sigma_{\ln \lambda^{ob}}^2}{B_\lambda} + \kappa(1+z)^2
\end{equation}
where the second term accounts for the redshift-dependent projection effect that is associated with redshift errors and the algorithm employed by the cluster finder \cite{Zhang2023}. 

It should be noted that, Costanzi et al. \cite{Costanzi2021} use $M_{500c}$, defined as the cluster mass enclosed within a radius with interior mean density of \textit{500} times the \textit{critical} density. 
So, when applying Eq.\eqref{eq:MRR}, we convert mass from $M_{200m}$ to $M_{500c}$ assuming that the halo density profile is described by the Navarro-Frenk-White (NFW) model \cite{NFW1997} with a concentration parameter from \cite{Duffy2008}.

Let $\Delta \ln M^{ob}_i$ denote the $i$th observed mass bin, and $\Delta z_j^{ob}$ denote the $j$th observed redshift bin. Finally, the expectation value of the number of detected galaxy clusters $N_{ij}$ is given by:
\begin{align}\label{eq:N2}
	N_{ij} 
	& \equiv N (\Delta \ln M^{ob}_i, \Delta z_j^{ob})  \notag
	\\
	& = \int_{\Delta \ln M^{ob}_i} d	\ln M^{ob} \int_{\Delta z_j^{ob}} d z^{ob}  \notag
	\\
	&\int d z \Delta \Omega \frac{dV}{d z d \Omega} ( z)  P( z^{ob} | z)
	\\
	& \int d \ln  M \frac{dn}{d \ln M}  (M,  z) P (\ln M^{ob} | \ln M,  z). \notag
\end{align}

\subsection{Cluster power spectrum}

The cluster power spectrum $P_c$ can be calculated from the matter power spectrum $P_m$ \cite{Wang2004}:
\begin{align}\label{eq:Pc}
		P_c (k^{\text{fid}}_{\perp}, k^{\text{fid}}_{\|},  z^{ob}) 
		& = b_\text{eff}^2 ( z) P_m (k^{\text{true}},  z)  \notag
		\\
		&\cdot \mathcal{R} (k^{\text{true}}_{\perp}, k^{\text{true}}_{\|}, z) \mathcal{AP} ( z) \mathcal{Z} ( z),
\end{align}
with a bias factor $b_\text{eff}(z)$ and several other factors, accounting for the redshift-space distortion (RSD) effect $\mathcal{R} (k^{\text{true}}_{\perp}, k^{\text{true}}_{\|})$, Alcock-Paczynski (AP) effect $\mathcal{AP} (z)$, and redshift uncertainty $\mathcal{Z} ( z)$. 
The wave numbers here are distinguished as fiducial ones $k^{\text{fid}}$ and true ones $k^{\text{true}}$, due to the AP effect.
Each term is detailed below.

The effective bias $b_{\text{eff}} (z)$ is an average of individual cluster bias $ b(M,z)$ at redshift $z$:
\begin{gather}
	b_\text{eff}(z) =
 \left[   \int d	\ln M^{ob}     \int d \ln M \frac{dn(M,  z)}{d \ln M} P (\ln M^{ob} | \ln M,  z)   \right]^{-1} 
\notag \\
	\int d	\ln M^{ob}    \int d\ln M \frac{dn(M,  z)}{d \ln M} P (\ln M^{ob} | \ln M,  z)    
     \cdot b(M, z ),\label{eq:beff}
\end{gather}
where the mass measurement uncertainty has also been considered.

The RSD effect refers to the distortion of the observed clustering pattern of clusters in redshift space due to their line-of-sight velocities, thus has an explicit dependence on the line-of-sight wave number $k^{\text{true}}_{\|}$:
\begin{equation}\label{eq:rsd}
	\mathcal{R} (k^{\text{true}}_{\perp}, k^{\text{true}}_{\|}, z) = \left[ 1 +\beta ( z) \left( \frac{k^{\text{true}}_{\|}}{k^{\text{true}}} \right)^2 \right]^2,
\end{equation}
in which, the redshift-distortion parameter is:
\begin{equation}\label{eq:beta}
	\beta ( z) \equiv \frac{1}{b_{\text{eff}} ( z)}  \frac{d \ln D_{\text{grow}} ( z)}{d \ln a},
\end{equation}
where $D_\text{grow} $ is the linear growth rate, and $a$ is the scale factor normalized to unity today.

The AP effect arises from assuming a fiducial cosmology to infer the position of a cluster from its observed redshift and angular position. If the "true" background cosmology differs from the fiducial cosmology, the inferred comoving radial and transverse distances are distorted with respect to the true ones. We fix the fiducial wave numbers $k^{\text{fid}}$, then compute the true wave numbers $k^{\text{true}}$:
\begin{align}\label{eq:APk}	
		k_{\perp}^{\text{true}} &= k^{\text{fid}}_{\perp} \cdot    D_A^{\text{fid}} ( z)   /   D_A ( z)    ,\notag
		\\
		k_{\|}^{\text{true}}  &= k^{\text{fid}}_{\|} \cdot    H( z)   /  H^{\text{fid}} ( z)   ,
\end{align}
and the AP effect term:
\begin{equation}\label{eq:AP}
	\mathcal{AP} ( z) = \left( \frac{H ( z)}{H^{\text{fid}} ( z)} \right) \left(
	\frac{D_A^{\text{fid}} ( z)}{D_A ( z)} \right)^2 ,
\end{equation}
where $H,D_A$ are the Hubble parameter and the comoving angular diameter distance, respectively. The superscripts $\{\text{fid},\text{true}\}$ represent fiducial and true models, respectively.

The treatment to account for redshift uncertainty in the cluster power spectrum is different from that in the number counts. The cluster distribution is 'smeared' along the radial direction, which damps the power spectrum by \cite{Blake2005}:
\begin{equation}\label{Pcz}
	\mathcal{Z} ( z) = 
	\exp \left\{
					- \left[ 
								\frac{k^{\text{true}}_{\|} }{H(z)}\sigma_{ z^{ob} \mid  z}( z) 
					\right]^2 	
				\right\},
\end{equation}
where $\sigma_{ z^{ob} \mid  z}( z)$ is the cluster redshift uncertainty Eq.\eqref{eq:sigmaz}.

\subsection{Neutrino prescription} \label{sec:prescription}

Galaxy clusters are non-linear objects that form at the peaks of the initial density field. 
The Press-Schechter (PS) theory\cite{PS1974} firstly provides a universal formulation for describing the HMF Eq.\eqref{eq:hmf}: $f(\nu)=\sqrt{\frac{2}{\pi}}\nu e^{-\nu^2/2}$. Here, $\nu\equiv \delta_\text{crit}/\sigma(M,z)$ specifies the peak height, a measure of the rarity of halo, and $\delta_\text{crit}=1.686$ is the critical density for halo collapse, weakly dependent on the underlying cosmological model.
Universality means that cosmological and redshift dependences enter the equation only through the peak height $\nu$, without affecting the functional form \cite{PS1974, ST1999, Jenkins2001}. 
However, accurate N-body simulations have revealed deviations from universality in redshift \cite{Tinker2008, Crocce2010, Bhattacharya2011} and in cosmological model \cite{Bhattacharya2011}.

Regarding cosmological models incorporating massive neutrinos, \cite{Villaescusa2014, Castorina2014, Costanzi2013} find that the impact of massive neutrinos can be minimized by replacing the \textit{total matter} density $ \rho_m=\rho_{cdm+b+\nu}$ in Eq.\eqref{eq:hmf} and Eq.\eqref{eq:MR} with the \textit{cold dark matter + baryon} density $ \rho_{cdm+b}$, \cite{Brandbyge2010, Marulli2011, Villaescusa2013}, and replacing the \textit{total matter} power spectrum $ P_m=P_{cdm+b+\nu}$  in Eq.\eqref{eq:sigmaR} with the \textit{cold dark matter + baryon} power spectrum $ P_{cdm+b}$. The physical reason behind this is that the clustering of neutrinos around the CDM overdensity is minimal and is unlikely to impact nonlinear collapse \cite{Ichiki2012, LoVerde2014_clustering}.
By employing this CDM prescription, \citep{Costanzi2013} shows that for simulations with massive neutrinos the HMF is well reproduced by the fitting formula of Tinker et al. \cite{Tinker2008}, which we will utilize in this paper.

According to the peak-background split (PBS) approach \citep{Bardeen1986}, the bias between the halo and matter overdensities $b(M,z)$ in Eq.\eqref{eq:beff} can be derived from the HMF. Therefore, the prescription with respect to the bias for cosmologies with massive neutrinos remains the same, i.e. replacing $P_m$ with $ P_{cdm+b}$, and $ \rho_{m}$ with $ \rho_{cdm+b}$,
which can recover the halo bias on large scales for simulations with massive neutrinos \cite{Villaescusa2014, Castorina2014}. We follow \cite{Villaescusa2014} and continue to use the fitting formula proposed by Tinker et al. \cite{Tinker2010} for halo bias.

However, this prescription only partially accounts for the effects of massive neutrinos on the large-scale bias. Massive neutrinos also introduce scale dependence in the bias due to two effects, both of which have been observed in N-body simulations: (i) halos trace CDM fluctuations rather than total (CDM + neutrino) matter fluctuations \cite{Villaescusa2014, Castorina2014}, and (ii) massive neutrinos introduce scale-dependence in the growth of CDM density perturbations \cite{Chiang2018, Chiang2019}.
The former can be eliminated by excluding neutrinos when defining bias, i.e. replacing $P_m$ in Eq.\eqref{eq:Pc} with $ P_{cdm+b}$. 
While, the latter, called the growth-induced scale-dependent bias (GISDB) \cite{Xu2021}, requires careful treatment.
To incorporate the GISDB effect, we employ a fitting formula from Eq.(49) of \cite{Munoz2018}, which is based on the spherical collapse model and the PBS approach \cite{LoVerde2014_bias, Munoz2018}. This method has been shown to be consistent with simulations \cite{Chiang2018}. The scale-dependence obtained by \cite{Munoz2018} is almost independent of mass and redshift\footnote{The scale-dependence does show weak dependence on redshift, which is neglected in our calculation. We agree a more accurate treatment would be using Relicfast directly rather than the fitting formula of Eq.~(\ref{eq:Rk}).}: 
\begin{align}\label{eq:Rk}
	\frac{b^L (k)}{b^L (k_{\text{ref}})} 
	& = [1 + \Delta_{\Lambda CDM} \tanh (\alpha k / k_{\text{eq}})] 
	\notag \\
	&\cdot  \left[ 1 + \frac{\Delta_L}{2}  \left( \tanh \left[ \frac{\log (q)}{\Delta_q} \right] + 1 \right) \right] 
	\notag  \\
	& \equiv R (k),
\end{align}
where $\Delta_{\Lambda CDM} = 4.8 \times 10^{-3}$, $\alpha = 4$, $\Delta_q = 1.6$, $\Delta_L = 0.55 f_{\nu}$ with $f_{\nu} \equiv \Omega_{\nu}/\Omega_m$ representing the neutrino fraction, $q = 5 k / k_{\text{fs}}$ (related to the free streaming scale Eq.\eqref{eq:kfs}). 
Here, $k_{\text{eq}}$ represents the scale of matter-radiation equality \cite{Planck2018}\footnote{Here, for simplicity, we treat $k_{eq}$ as a constant.}, 
and $k_{\text{ref}}$ is a reference wave number on large scales, which we set to be $10^{-4} \hMpc$. 
The superscript $L$ denotes Lagrangian bias. Since we use $P_{cdm+b}$ to calculate the cluster power spectrum, the Eulerian bias is given by $b = 1 + b^L$. Finally, we can replace the scale-independent effective bias $b_{\text{eff}}(z) = b_{\text{eff}}(k_{\text{ref}}, z)$ in Eq.\eqref{eq:beff} by the scale-dependent effective bias:
\begin{align}
	b_{\text{eff}} (z) \rightarrow b_{\text{eff}} (k, z) = [b_{\text{eff}} (k_{\text{ref}},z) - 1] \cdot R (k) + 1.
\end{align}
Ignoring these two neutrino-induced scale dependences in the bias, will have an impact on cosmological parameter inference in galaxy survey \cite{LoVerde2016, Raccanelli2018, Vagnozzi2018, Xu2021}. In this paper, we focus on the latter effect, i.e. GISDB effect in cluster survey.

\subsection{Fisher matrix}

We employ the Fisher matrix techniques to assess the constraining power of clusters. It is defined as the ensemble average of the second derivatives of the logarithm of likelihood function with respect to the parameters $(p_\mu, p_\nu)$:
\begin{equation}
	\mathcal{F}_{\mu\nu} \equiv - 	\left<     \frac{\partial^2 \ln \mathcal{L}}{\partial p_\mu \partial p_\nu}  \right>,
\end{equation}
where $\mathcal{L}$ represents the likelihood function. The inverse of the Fisher matrix, denoted as $\mathcal{F}^{-1}$, provides the best possible covariance matrix for the measurement errors on the parameters.

Assuming the observed cluster number count $D$ obeys a Poisson distribution with an expected value and variance $N$ \cite{Holder2001}, the logarithm of the likelihood function $\mathcal{L}(\bm D)$ can be expressed as:
\begin{equation}
	\ln \mathcal{L}(\bm D)= \sum_{ij} D_{ij} \ln N_{ij} - N_{ij} - \ln D_{ij}!,
\end{equation}
where the subscripts ${i,j}$ represent observed mass and redshift bins, $N_{ij}$ denotes the expected number count in the fiducial model in bin $(i,j)$ Eq.\eqref{eq:N2}. 
The Fisher matrix for the cluster number count $N$, denoted as $\mathcal{F^N_{\mu \nu}}$, is then given by:
\begin{equation}\label{eq:FN}
	\mathcal{F}_{\mu \nu}^N = \sum_{i, j} \frac{\partial N_{ij}}{\partial p_{\mu}}  \frac{\partial N_{ij}}{\partial p_{\nu}} 
	\frac{1}{ N_{ij}}.
\end{equation}

For the observed cluster power spectrum, assuming it follows a Gaussian distribution with a mean value $P_c$ Eq.\eqref{eq:Pc} and a fractional variance $\sigma_P^2/P_c^2=2/\left(V_k V_\text{eff}\right)$ \cite{Wang2004, Wang2005}, the Fisher matrix, denoted as $\mathcal{F}_{\mu \nu}^P$, can be expressed as:
\begin{equation}\label{eq:FPc}
	\mathcal{F}_{\mu \nu}^P =
	\sum_{i, j} 
	\frac{\partial \ln  P_{c,ij}}  {\partial p_{\mu}} 
	\frac{\partial \ln  P_{c,ij}}  {\partial p_{\nu}} 
	\frac{\left( V_k V_{\mathrm{eff}}	\right)_{ij}}{2},
\end{equation}
where the subscripts $\{i,j\}$ represent the two-dimensional $\vec{k}^\text{fid}$-space cells and the observed redshift bins, respectively.

$V_{\text{eff}}$ represents the effective volume probed by the survey \cite{FKP1994, Hu2003}:
\begin{align}\label{eq:Veff}
&	V_{\mathrm{eff}} (k^{\text{fid}}_{\perp}, k^{\text{fid}}_{\|}, z^{ob}) 
	= \int dV_s
	\left[ \frac{ P_c }{\mathcal{AP}\bar{n}^{-1} + P_c} \right]^2  \notag
\\
&
	= \int_{\Delta  z^{ob}}   d z^{ob} \Delta \Omega \frac{dV^{ob}}{d z^{ob} d \Omega}
\\
&
 \left[ \frac{ P_c (k^{\text{fid}}_{\perp},k^{\text{fid}}_{\|},  z^{ob})}
  {\mathcal{AP}(z^{ob})\bar{n}^{-1} ( z^{ob}) + P_c (k^{\text{fid}}_{\perp}, k^{\text{fid}}_{\|},  z^{ob})} \right]^2.
\end{align}
The impact of redshift uncertainty on the volume factor is negligible. For simplicity in calculations, we have disregarded it, i.e. $dV^{ob}/(d z^{ob} d \Omega)=dV/(d z d \Omega)$ without a integration over $dz P( z^{ob} | z) $. However, integration over $dz P(z^{ob} | z)$ is necessary only for $\bar{n}$, the expected average number density (similar to Eq.\eqref{eq:N2}):
\begin{equation}\label{eq:barn}
	\begin{array}{ll}
		\bar{n}(z^{ob})
		& = \int d	\ln M^{ob} \int d zP( z^{ob} | z)\\
		\\
		& \int d \ln M \frac{dn}{d\ln M}  (M,  z) P (\ln M^{ob} | \ln M,  z),
	\end{array}
\end{equation}
and $V_k$ represents the cylindrical volume factor in $k$-space:
\begin{equation}\label{eq:Vk}
	V_k (k^{\text{fid}}_{\perp}, k^{\text{fid}}_{\|}) = \frac{2 \pi \Delta
		(k^{\text{fid}}_{\perp})^2 \Delta k^{\text{fid}}_{\|}}{(2 \pi)^3}.
\end{equation}

\subsection{Parameters}\label{sec:params}

\begin{table}
	\caption{\label{fiducial}
	Parameters and their fiducial values.	
	}
	\begin{ruledtabular}
		\begin{tabular}{c|c|c}
			Parameter & Description & Fiducial values \\
        \hline
        \multicolumn{3}{c}{Cosmological parameters} \\
        \hline
			$ \Omega_bh^2$			 & Baryon density & 0.02242  \\
			$ \Omega_mh^2$		 & Matter density \footnote{including massive neutrinos}& 0.14240  \\
			$ \Omega_{\nu}h^2$		 & Massive neutrinos density & 0.0006442\\
			$ \Omega_{DE}$		 &	DE density & 0.6889\\
			$ w_0$ 							& DE EoS parameter & -1 \\
			$ w_a$ 						& DE EoS parameter & 0  \\
			$ n_s$ 						& Spectral index & 0.9665 \\
			$ \sigma_8$					 & Normalization of perturbations & 0.8102  \\
			\hline
    	\multicolumn{3}{c}{Scaling relation parameters} \\
        \hline
			$ A_\lambda $ 					& Amplitude in $\left< \ln \lambda^{ob} \right>$  
            \footnote{see Eq.\eqref{eq:MRR}\label{fn:MRR}} & 79.8 \\
			$ B_\lambda $ & Mass dependence in $\left< \ln \lambda ^{ob} \right>$ \footref{fn:MRR}& 0.93 \\
			$ C_\lambda $					 & Redshift dependence in $\left< \ln \lambda ^{ob} \right>$ \footref{fn:MRR}& -0.49 \\
			$ D_\lambda $		 & Intrinsic scatter in $\sigma_{\ln \lambda ^{ob}}$ \footnote{see Eq.\eqref{eq:sigmalambda}}& 0.217  \\

			$ \kappa $		 & Projection effect parameter \footnote{see Eq.\eqref{eq:sigmaMob}\label{fn:sigmaMob}}& 0.01   \\
            $ B_{M0}$ 					& Constant term in $M_\text{bias}$ \footnote{see Eq.\eqref{eq:Mbias} \label{fn:Mbias} }& 0 \\
			$ \alpha $ 			& Redshift dependence in $M_\text{bias}$ \footref{fn:Mbias} & 0 \\
        \hline
        \multicolumn{3}{c}{Derived parameter} \\
        \hline
			$ M_{\nu}(eV) $ & Total neutrino mass &  0.06
		\end{tabular}
	\end{ruledtabular}
\end{table}

The basic cosmological model we consider is an extension of the flat $ \Lambda $CDM model by adding one species of massive neutrino, labeled as $\nu \Lambda $CDM model, with the other two species assumed to be massless. We also consider $\nu w_0w_a $CDM model, incorporating dynamical dark energy (DE) with an equation of state (EoS) represented by $w(a) = w_0 + (1-a)w_a$ \cite{Chevallier_Polarski2001, Linder2003}.
The set of cosmological parameters encompasses:
\begin{align}
	\{ \Omega_b h^2, \Omega_m h^2, \Omega_{\nu} h^2, \Omega_{DE}, w_0, w_a, n_s, \sigma_8 \},
\end{align}
 which denotes the baryon density, matter density, neutrino density, DE density, DE EoS parameters, scalar power spectrum index and present-day normalization of the primordial power spectrum, respectively.
 The fiducial values for $\{ \Omega_b h^2, \Omega_m h^2, \Omega_{DE}, n_s, \sigma_8 \}$ follows \cite{Planck2018}, while $\Omega_{\nu} h^2=0.0006442$ corresponding to $M_{\nu}=0.06 $ eV through \cite{Lesgourgues2006}:
\begin{align}\label{eq:9314}
	 \Omega_{\nu}h^2=\frac{M_\nu}{93.14 eV},
\end{align}
and fiducial values for DE EoS parameters are set as $w_0=-1,w_a=0$.

Additionally, there are a set of parameters associated with the scaling relation (SR):
\begin{align}
	\{ A_\lambda, B_\lambda , C_\lambda, D_\lambda, \kappa, B_{M0}, \alpha\},
\end{align}
where the first four parameters $\{ A_\lambda, B_\lambda , C_\lambda, D_\lambda \}$ pertain to the mass-richness scaling in Eq.\eqref{eq:MRR} and Eq.\eqref{eq:sigmalambda}, and their fiducial values follow Table IV and Eq.(B1) of \cite{Costanzi2021}: \{ 79.8, 0.93, -0.49, 0.217\}. 
For the projection effect term in Eq.\eqref{eq:sigmaMob}, we set $\kappa=0.1^2$ accounting for a possible contamination rate of about 10\% at low redshift for massive clusters for CSST \cite{Zhang2023, Wen2009}. The mass bias parameters in Eq.\eqref{eq:Mbias} are specified as $B_{M0} = 0$ and $\alpha = 0$.
	
All parameters and their fiducial values are summarized in Table \ref{fiducial}.

CSST will perform a wide-field imaging survey in the $NUV$, $u$, $g$, $r$, $i$, $z$ and $y$ bands and a slitless spectroscopic survey in the $GU$, $GV$ and $GI$ bands simultaneously with an area of 17,500 deg$^2$, which corresponds to $\Delta \Omega=5.33$ sr.
The CSST camera has a field of view of 1.1 deg$^2$. The imaging survey reaches a depth of $i=25.9$ for 5\,$\sigma$ point source detection with a spatial resolution of $\sim 0.15$ arcsec. The slitless spectroscopic survey reaches a depth of $GI=23.2$ for point sources with a spectral resolution $R\ge200$. The magnitude limit for extended sources is about 1 mag shallower \cite{Zhan2021,Gong2025}.

We assume the CSST cluster redshift uncertainty to be $\sigma_z=0.001$ (as in Eq.\eqref{eq:sigmaz}). This is reasonable for the CSST slitless grating spectroscopic data \cite{Zhou2024}, which are used to determine spectroscopic redshifts of clusters up to near $z=1.5$ \citep{Gong2019}. We expect that only a small fraction of clusters has their redshifts determined only by photometric redshifts of galaxies with an uncertainty of $\sim0.02$ \cite{Zhou2022}. The redshift uncertainty of each cluster as a whole would be reduced after averaging the redshifts of the member galaxies. Thus, we adopt an optimistic value of $0.001$.
We assume that the cluster detection is reliable up to $z_{\max}=1.5$, at which the 4000\AA~break is redshifted to $y$ band, using either redshift-based method \cite{Wen2021, Yang2021} or color-based method \cite{Rykoff2014}.
Here, we take the cluster catalog constructed from DESI imaging surveys \cite{Wen2024} as a reference for CSST cluster catalog. If we set the observed mass threshold of $M_{500c} \geq 0.7 \times 10^{14} \Msun$, equivalent to $M_{200m} \geq 0.836 \times 10^{14} \hMsun$, the DESI cluster catalog has a completeness of $\geq 90\%$ and a purity of $\geq 90\%$. Considering that the CSST provides deeper data with more bands and slitless spectroscopy, this is a conservative choice for the CSST cluster catalog.

For cluster number counts, the observed mass bin size is set to be $\Delta \ln M^{ob} [\Msun]$ = 0.2, and the observed redshift bin size is $\Delta z^{ob}=0.05$. While for the cluster power spectrum, we choose $\Delta z^{ob}=0.2$, ensuring adequate number of clusters within each redshift bin. The fiducial wavenumber for the CSST cluster power spectrum spans from $k^{\text{fid}}_{\perp, \|} = 0.005$ to $0.15$ \Mpcc \cite{Hu2003, Wang2005,FH07}, corresponding roughly to the largest observable scale in the transverse or los direction for a given redshift bin and the nonlinear scale at z=0 (see e.g. Fig. 8.3 in \cite{Dodelson2020}), respectively, and we choose a bin size of $\Delta k^{\text{fid}}_{\perp,	\|} = 0.005$ \Mpcc.

By summing over the aforementioned bins, we obtain the Fisher matrices $\mathcal{F}_{\mu \nu}^N$ for the cluster number counts Eq.\eqref{eq:FN} and $\mathcal{F}_{\mu \nu}^P$ for power spectrum Eq.\eqref{eq:FPc}.
Finally, we can calculate the $1\,\sigma$ marginalized error on a parameter $p_{\mu}$ as $\sigma(p_{\mu})=\sqrt{(\mathcal{F}^{-1})_{\mu \mu}}$, as shown in \autoref{tab:result}.

\section{results}\label{sec:results}
\begin{figure}
	\includegraphics[width=0.9\linewidth]{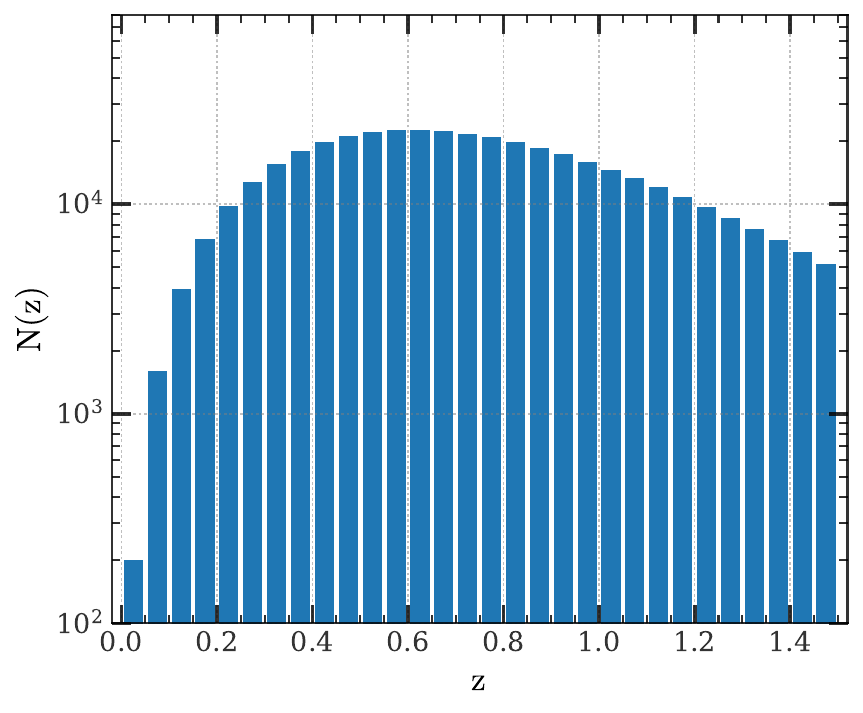}
	\caption{Cluster number counts as a function of redshift $z$. }
	\label{fig:abundance}
\end{figure}

\begin{table*}
	\caption{\label{tab:result}
     Marginalized $1\,\sigma$ errors forecasted for all the parameters described in \autoref{sec:params}. When a parameter is kept fixed, we use '–'.
    The header specifies the different cosmological models ($\nu \Lambda $CDM and $\nu w_0w_a $CDM), scenarios for our knowledge of the scaling relation (self-calibration and known SR), and probes (cluster number counts $N$, cluster power spectrum $P$, and their combination $N+P$).There are almost no constraints from $P$ on the SR parameters, denoted by '/', and adding priors to these parameters does not affect the main results. For complete description, please refer to the main text.
    }
	\begin{ruledtabular}
		\begin{tabular}{c|cccccc|cccccc} 
         
        model & \multicolumn{6}{c}{$\nu \Lambda $CDM} &  \multicolumn{6}{c}{$\nu w_0 w_a $CDM}  \\ 
         \hline
         scenario & \multicolumn{3}{c}{Self-calibration} & \multicolumn{3}{c}{Known SR} &  \multicolumn{3}{c}{Self-calibration} &  \multicolumn{3}{c}{Known SR} \\ 
         \hline
        probe & $N$     & $P$     & $N+P$   & $N$     & $P$     & $N+P$   & $N$     & $P$     & $N+P$   & $N$     & $P$     & $N+P$   \\
        \hline
        $\Omega_b h^2$& 1.831 & 0.006 & 0.005      & 0.454 & 0.004 & 0.004      & 1.968 & 0.008 & 0.005      & 0.468 & 0.005 & 0.004      \\
        $\Omega_m h^2$& 5.741 & 0.026 & 0.022      & 1.379 & 0.018 & 0.015      & 6.200 & 0.039 & 0.023      & 1.412 & 0.022 & 0.015      \\
        $\Omega_{\nu} h^2$& 0.056 & 0.002 & 0.001      & 0.008 & 0.002 & 3.4e-04      & 0.059 & 0.003 & 0.001      & 0.012 & 0.002 & 3.6e-04      \\
        $\Omega_{DE}$& 0.122 & 0.014 & 0.008      & 0.004 & 0.008 & 0.001      & 0.161 & 0.028 & 0.010      & 0.017 & 0.017 & 0.004      \\
        $w_0$& - & - & -      & - & - & -      & 0.083 & 0.168 & 0.057      & 0.067 & 0.130 & 0.050      \\
        $w_a$& - & - & -      & - & - & -      & 1.158 & 0.664 & 0.225      & 0.463 & 0.483 & 0.191      \\
        $n_s$& 3.790 & 0.046 & 0.039      & 0.827 & 0.028 & 0.017      & 4.177 & 0.069 & 0.040      & 0.836 & 0.035 & 0.017      \\
        $\sigma_8$& 0.258 & 0.018 & 0.010      & 0.010 & 0.004 & 0.002      & 0.269 & 0.036 & 0.010      & 0.058 & 0.006 & 0.003      \\
        \hline
        $A_\lambda$& 1.5e+02 & / & 1.1e+01      & - & - & -      & 2.0e+02 & / & 1.2e+01      & - & - & -      \\
        $B_\lambda$& 1.024 & / & 0.203      & - & - & -      & 1.356 & / & 0.207      & - & - & -      \\
        $C_\lambda$& 1.628 & / & 0.328      & - & - & -      & 2.294 & / & 0.336      & - & - & -      \\
        $D_\lambda$& 0.363 & / & 0.074      & - & - & -      & 0.415 & / & 0.075      & - & - & -      \\
        $\kappa$& 0.466 & / & 0.129      & - & - & -      & 0.539 & / & 0.136      & - & - & -      \\
        $B_{M0}$& 0.860 & / & 0.029      & - & - & -      & 1.006 & / & 0.033      & - & - & -      \\
        $\alpha$& 0.431 & / & 0.045      & - & - & -      & 0.434 & / & 0.047      & - & - & -      \\
        \hline
        $M_{\nu}(eV)$& 5.215 & 0.201 & 0.117      & 0.706 & 0.194 & 0.031      & 5.458 & 0.293 & 0.127      & 1.095 & 0.223 & 0.034      \\
        \end{tabular}
	\end{ruledtabular}
\end{table*}

In this section, we present the main results forecasted for the CSST. The redshift distribution of the CSST clusters is shown in Figure \ref{fig:abundance}, with a total number count of 408,037. (This is slightly different from our previous prediction of 415,000 in \cite{Zhang2023} due to the slight difference in the fiducial parameters.)
The predicted $1\,\sigma$ marginalized errors for all parameters, especially for the total neutrino mass $M_\nu$, are shown in \autoref{tab:result}, for different probes, cosmological models, and scenarios.
Specifically, these constraints are derived from the cluster number counts $N$, the cluster power spectrum $P$, and their combination $N+P$, and for the two cosmological models: the $\nu \Lambda $CDM model and the $\nu w_0w_a$CDM model.
To address the impact of the major systematic in cluster cosmology, i.e. the uncertainty in the SR, we consider two scenarios: (1) the self-calibration scenario, where the SR parameters are jointly constrained with the cosmological parameters; and (2) the known SR scenario, where the SR parameters are fixed.

For clarity, the constraints on $M_{\nu}$ obtained from $N$, $P$ and $N+P$ are denoted as $\sigma^{N}(M_{\nu})$, $\sigma^{P}(M_{\nu})$ and $\sigma^{N+P}(M_{\nu})$ respectively.

\subsection{$\nu \Lambda $CDM model} \label{sec31}

\begin{figure*}
	\centering
	\includegraphics[width=0.9\linewidth]{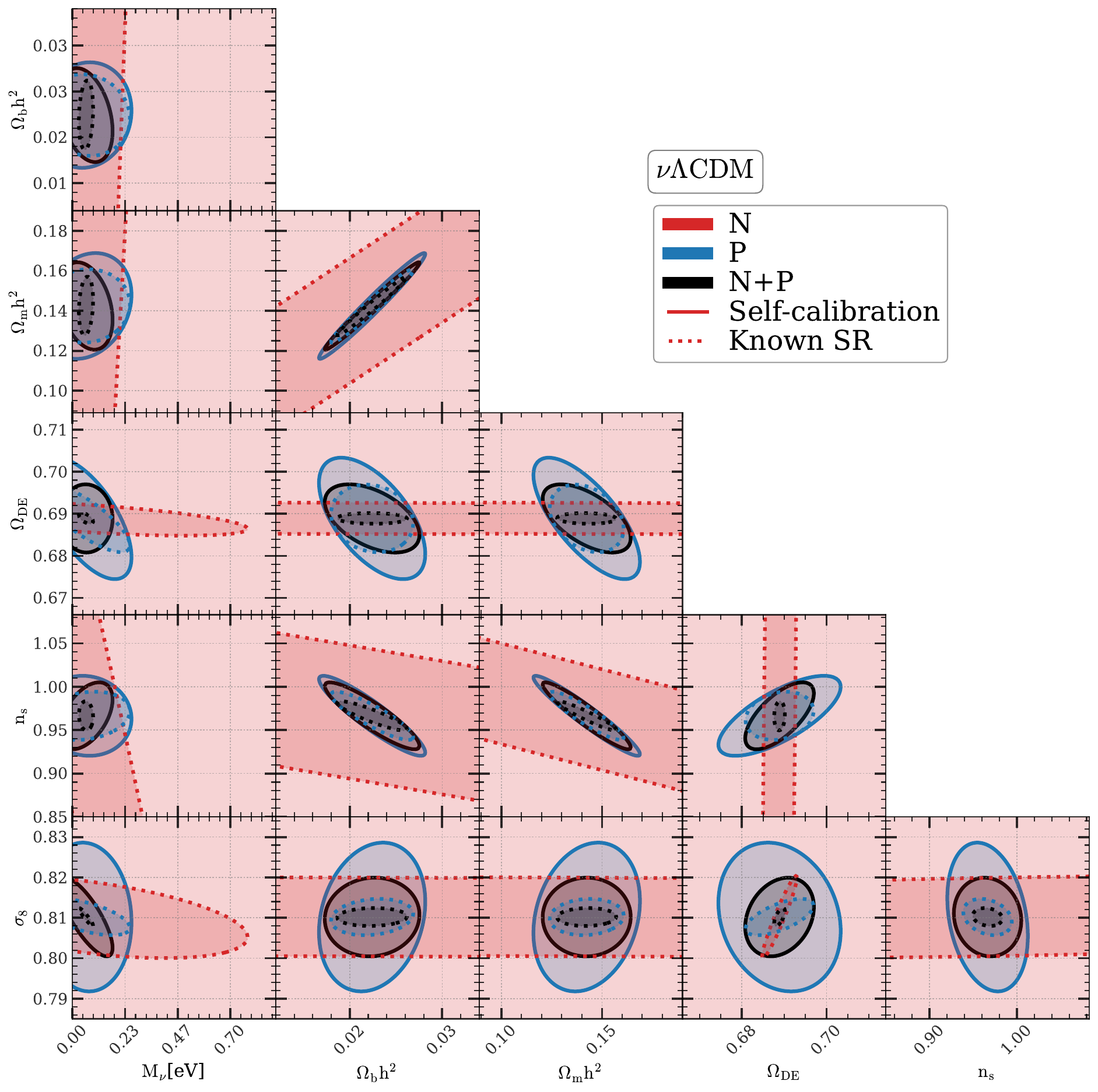}
	\caption{ Marginalized constraints on cosmological parameters for the $\nu \Lambda $CDM model. 
    All contours show the 68\% confidence regions.
    Different colors (red, blue and black) represent different probes: the cluster number counts $N$, cluster power spectrum $P$, and their combination $N+P$ in turn.
    Solid and dotted lines represent the self-calibration and known SR scenarios respectively.
	}
	\label{fig:ContourNuLCDM}
\end{figure*}

We begin with the $\nu \Lambda $CDM model, whose constraints are shown in columns 2-7 of \autoref{tab:result}.
The corresponding contour plot is shown in \autoref{fig:ContourNuLCDM} for the cosmological parameters $\{ M_\nu,\Omega_b h^2, \Omega_m h^2, \Omega_{DE}, n_s, \sigma_8 \}$.

It is evident that constraints on all cosmological parameters primarily come from $P$, except for $\Omega_{DE}$ in the known SR scenario. 
This is because $N$ is sensitive to the integration of the $\textit{cdm+b}$ power spectrum, rather than its shape, resulting in weaker constraints on $\Omega_b h^2$, $n_s$ and $M_\nu$ \cite{Wang2004, Wang2005, Carbone2012}.
Consequently, the constraint on $M_{\nu}$ from $N$ primarily stems from the influence of neutrinos on the amplitude of the $\textit{cdm+b}$ power spectrum, leading to a strong degeneracy between $M_{\nu}$ and $\sigma_8$. 
In summary, the constraint on $M_{\nu}$ from $N$ is very weak, $\sigma^{N} (M_{\nu})=5.215\ (0.706)$ eV in the self-calibration (known SR) scenario.

The power spectrum $P$ proves to be a more sensitive probe of the neutrino mass compared to $N$, providing a constraint of $\sigma^{P} (M_{\nu})=0.201\ (0.194)$ eV on its own. When combined with $N$, the improvement factor in the self-calibration (known SR) scenario is 1.72 (6.26).

The incorporation of SR information has a substantial impact on $N$, resulting in an enhancement of constraints on all cosmological parameters by a factor of at least four. Particularly, the constraint on $M_{\nu}$ experiences a significant improvement by a factor of 7.4, shifting from $5.215$ to $0.706$ eV.
In contrast, the influence of SR information on $P$ is less significant, which is reflected in the large errors of SR parameters in the third column, since the SR parameters only affect $b_{\text{eff}}$. Consequently, the improvement in the constraint on $M_{\nu}$ is only a factor of 1.04, from $0.201$ to $0.194$ eV, when incorporating SR information.
In the combined analysis, the inclusion of SR information strengthens the constraint from $0.117$ to $0.031$ eV, which corresponds to a $\sim 2\sigma$ detection of the minimum neutrino mass of $0.06$ eV.

\subsection{$\nu w_0 w_a $CDM model} \label{sec32}
\begin{figure}
	\centering
	\includegraphics[width=0.9\linewidth]{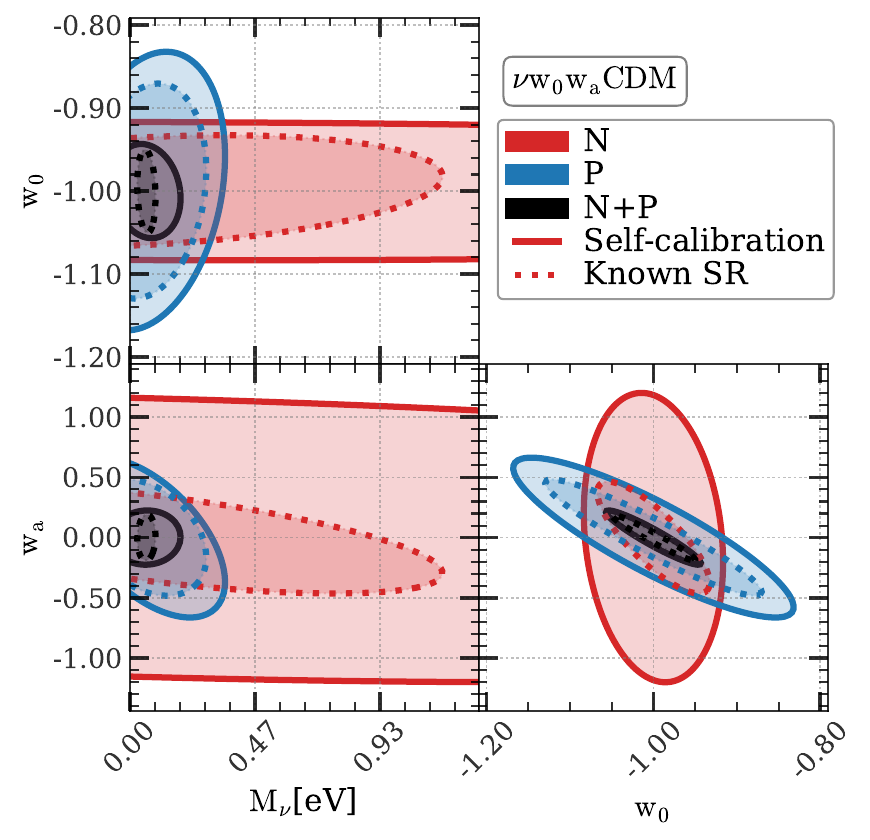}
	\caption{ Similar to \autoref{fig:ContourNuLCDM}, but for the $\nu w_0 w_a $CDM model.
	}
	\label{fig:ContourNuw0waCDM}
\end{figure}

Next we extend the $\nu \Lambda $CDM model to the $\nu w_0 w_a $CDM model in order to investigate the impact of dynamical dark energy on the neutrino constraint, results showing in columns 8-13 of \autoref{tab:result}.
The corresponding contour plot is shown in \autoref{fig:ContourNuw0waCDM} for the cosmological parameters $\{ M_\nu, w_0, w_a \}$ with the other parameters marginalized over.

It is expected that including the parameters $\{w_0, w_a\}$ will weaken the constraint on $M_\nu$ since both dark energy and massive neutrinos affect the growth rate of structures in the Universe.
As depicted in \autoref{fig:ContourNuw0waCDM}, there is a positive correlation between $M_\nu$ and $w_0$, as well as a negative correlation between $M_\nu$ and $w_a$, for both $N$ and $P$. However, the direction of degeneracy is different between $N$ and $P$, resulting in the constraint on $M_\nu$ from the combination $N+P$ being less affected by the presence of the dynamical dark energy \cite{Mak2013}.

 Here we consider the known SR scenario as an example. Adding $\{w_0, w_a\}$ results in a weakening of $\sigma (M_{\nu})$ by a factor of 1.55 and 1.15 for $N$ and $P$, respectively. In contrast, the combination $N+P$ shows a weakening factor of only 1.10. The same holds true for the self-calibration scenario, where the weakening factor for the combination $N+P$ is 1.08.

\bigskip
In summary, the total neutrino mass, $M_\nu$, is much better constrained by the cluster power spectrum than by the cluster number counts. The two probes are highly complementary in constraining $M_\nu$. The inclusion of dynamical dark energy has negligible impact on the joint neutrino mass constraint from $N+P$. Availability of perfect SR knowledge plays a significant role in constraining $M_\nu$.

\section{discussions}\label{sec:discussions}

\begin{table*}
	\caption{\label{tab:resultDis}
     The $1\,\sigma$ marginalized errors forecasted for the the total neutrino mass under different conditions.
    The 4$^{th}$ row give the fiducial results, as shown in \autoref{tab:result}.
    The following rows show the results under the conditions of: partial SR knowledge (5$^{th}$ row), $z_{max}=2$ (6$^{th}$ row), $k_{\text{max}}=0.07$ \Mpcc (7$^{th}$ row),
    $\sigma_z=0.01$ (8$^{th}$ row),  HMF uncertainty (9$^{th}$ row), bias uncertainty (10$^{th}$ row), and without considering the GISDB effect (11$^{th}$ row).
    The symbol `/' represents that the constraint is the same as the fiducial case.
    }
	\begin{ruledtabular}
		\begin{tabular}{c|cccccc|cccccc} 
        model & \multicolumn{6}{c}{$\nu \Lambda $CDM} &  \multicolumn{6}{c}{$\nu w_0 w_a $CDM}  \\ 
         \hline
         scenario & \multicolumn{3}{c}{Self-calibration} & \multicolumn{3}{c}{Known SR} &  \multicolumn{3}{c}{Self-calibration} &  \multicolumn{3}{c}{Known SR} \\ 
         \hline
        probe & $N$     & $P$     & $N+P$   & $N$     & $P$     & $N+P$   & $N$     & $P$     & $N+P$   & $N$     & $P$     & $N+P$   \\
        \hline
        fiducial & 5.215 & 0.201 & 0.117      & 0.706 & 0.194 & 0.031      & 5.458 & 0.293 & 0.127      & 1.095 & 0.223 & 0.034      \\
        \hline
        partial& 2.950 & 0.201 & 0.116& / & / & / & 4.590 & 0.292 & 0.125& / & / & / \\
        $z_{max}=2$& 3.072 & 0.196 & 0.104      & 0.642 & 0.190 & 0.030      & 3.329 & 0.287 & 0.118      & 0.909 & 0.218 & 0.033      \\
        $k_{\text{max}}=0.07$ \Mpcc& / & 0.319 & 0.187      & / & 0.303 & 0.056      & / & 0.467 & 0.248      & / & 0.355 & 0.060      \\

        $\sigma_z=0.01$& 5.266 & 0.514 & 0.188      & 0.706 & 0.364 & 0.065      & 5.509 & 0.653 & 0.241      & 1.098 & 0.475 & 0.073      \\

        HMF & 5.238 & 0.201 & 0.127      & 0.830 & 0.199 & 0.066      & 6.524 & 0.293 & 0.235      & 2.945 & 0.224 & 0.135      \\
        
        bias & / & 0.201 & 0.161      & / & 0.200 & 0.091      & / & 0.293 & 0.249      & / & 0.290 & 0.119      \\

        w/o GISDB& / & 0.187 & 0.140      & / & 0.158 & 0.064      & / & 0.277 & 0.191      & / & 0.229 & 0.075      \\
        
        \end{tabular}
	\end{ruledtabular}
\end{table*}

In this section, we study several effects that may change our forecasts, including knowledge on SR parameters, the maximum redshift, uncertainties in redshift measurements, HMF and bias, as well as the GISDB effect, with results presenting in \autoref{tab:resultDis}, and compare with literature at the end of this section.

\subsection{SR parameters}\label{sec:DisSR}
\begin{figure}
	\centering
	\includegraphics[width=0.9\linewidth]{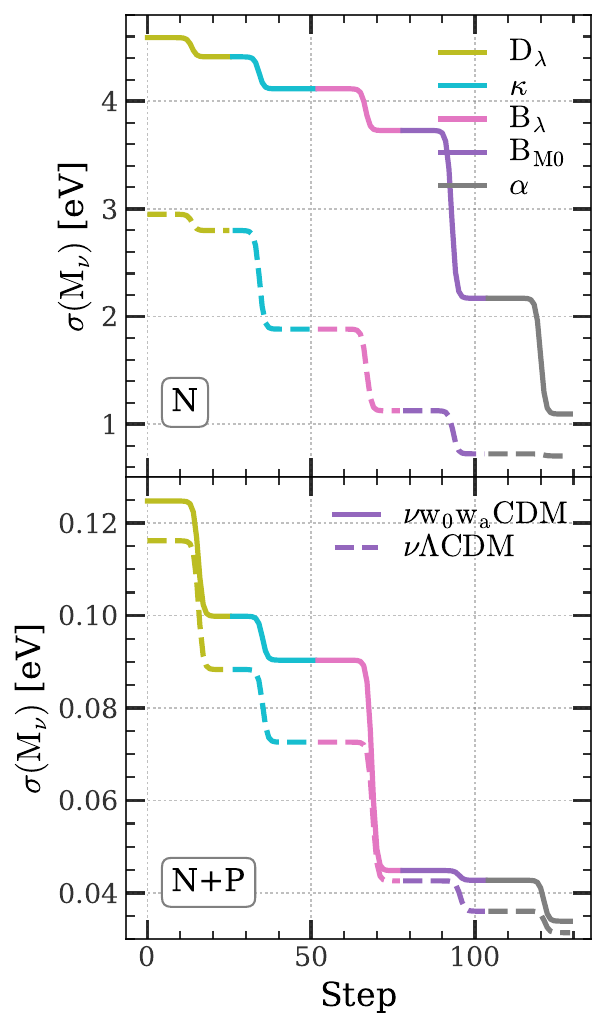}
	\caption{ The constraint on the total neutrino mass $\sigma(M_\nu)$ as we increase prior information on each SR parameter $\{B_\lambda, D_\lambda, \kappa, B_{M0}, \alpha\}$ denoted with different colors.
    Each step represents the addition of a prior in a logarithmic space, from $10^4$ to $10^{-4}$, which is wide enough to cover the 'totally ignorant' to 'totally known' cases.
    Solid and dashed lines represent $\nu w_0 w_a $CDM and $\nu \Lambda $CDM model respectively.
    The upper and the lower panel represents the constraint from $N$ and $N+P$ respectively.
	}
	\label{fig:PriorB0.5}
\end{figure}

In our analysis, we do not directly use the mass observable for CSST, i.e. the optical richness. 
Although the mass-richness relation (MRR) has been calibrated in numerous works (e.g. \cite{Simet2017, Murata2019, Capasso2019, Bleem2020, Costanzi2021}), it will be affected by survey specifics and cluster finders \cite{Old2014, Adam2019}. Therefore it is inappropriate to apply a MRR calibrated in one specific survey to another survey.
Instead, we employ a simple parametrization to account for our current lack of knowledge about the details of the mass-observable relations and their potential dependences on redshift and mass.
However, our approach is not completely devoid of information. 
Rather than treating $\sigma_{\ln M}$ as a nuisance parameter, we relate it with $\sigma_{\ln \lambda}$, and assume $\sigma_{\ln \lambda}^2$ to be composed of a constant intrinsic scatter term and a Poisson-like term.
In our default analysis, we have selected a MRR to model the Poisson-like term of $1/\exp \left<\ln \lambda^{ob}(M) \right>$. However, in real surveys, we can also use the richnesses for the detected clusters directly, then
there will be no need for an additional MRR to specify the Poisson-like term.
Consequently, we try fixing the parameters $\{ A_\lambda, B_\lambda, C_\lambda \}$ in Eq.\eqref{eq:MRR} while retaining $B_\lambda$ as a free parameter in Eq.\eqref{eq:sigmaMob} to describe this scenario, which we refer to as the partial self-calibration scenario.

The 5$^{th}$ row in \autoref{tab:resultDis} shows the result. Similar to the previous section, SR parameters have a greater impact on $N$ than on $P$. In the partial self-calibration scenario, the constraint on $M_\nu$ is improved by a factor of 1.8 (1.2) in the $\nu \Lambda $CDM ($\nu w_0 w_a $CDM) model for $N$. However, the improvement is minimal for the constraint from $P$ or $N+P$.

We proceed to investigate the effect of the remaining SR parameters, $\{B_\lambda, D_\lambda, \kappa, B_{M0}, \alpha\}$, on $\sigma(M_\nu)$ by adding the prior information, as depicted in \autoref{fig:PriorB0.5}.
The improvement trend is similar for both $\nu \Lambda $CDM and $\nu w_0 w_a $CDM model.
The upper panel illustrates that $\sigma^N(M_\nu)$ is primarily influenced by $B_{M0}$ and $\alpha$, which determine the total number counts.
On the other hand, the lower panel demonstrates that $\sigma^{N+P}(M_\nu)$ is more sensitive to the scatter parameters $B_\lambda, D_\lambda$ and $\kappa$, which suggests we would better select an observable with smaller scatter when using the combination probe $N+P$ to constrain the total neutrino mass.

Recently, we have calibrated the intrinsic mass-richness relation of clusters with THE THREE HUNDRED hydrodynamic simulations \cite{Chen2024}. Our findings indicate that employing a skewed Gaussian distribution to model the richness at a fixed halo mass results in a more accurate compared to the conventional use of a log-normal distribution. We plan to incorporate these findings in future cluster analyses.

\subsection{$z_{max}$}\label{sec:DisZmax}
\begin{figure}
	\centering
	\includegraphics[width=0.9\linewidth]{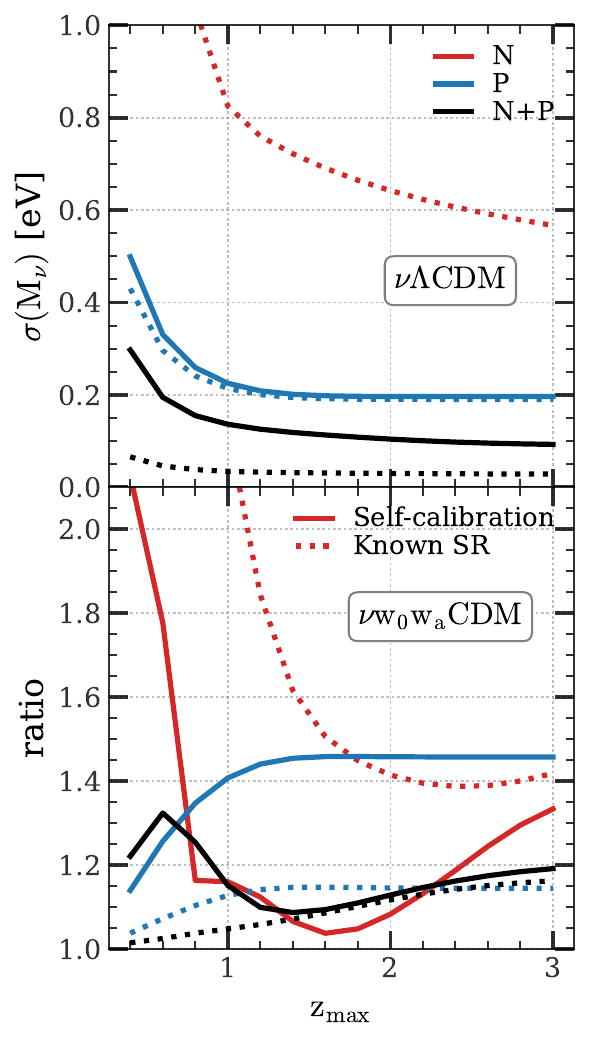}
	\caption{ 
    The impact of $z_{max}$ on $\sigma(M_\nu)$.
    Different colors (red, blue and black) represent different probes: the cluster number counts $N$, cluster power spectrum $P$, and their combination $N+P$.
    Solid and dotted lines represent self-calibration and known SR scenario respectively.
    The upper panel shows the $\sigma(M_\nu)$ in $\nu \Lambda $CDM model. 
    The lower panel shows the ratio of $\sigma(M_\nu)$ in $\nu w_0 w_a $CDM to $\sigma(M_\nu)$ in $\nu \Lambda $CDM model.
	}
	\label{fig:zmax}
\end{figure}

In our fiducial analysis, we assume a maximum redshift of $z_{\max}=1.5$ for cluster detection with CSST. However, by improving the cluster selection algorithm or combining data from other surveys like Euclid \cite{Sartoris2016}, it may be possible to extend $z_{\max}=1.5$ to be even higher such as to $z_{\max}=2$ \cite{Zhang2023}.
The 6$^{th}$ row in \autoref{tab:resultDis} shows the result with $z_{max}=2$. Compared to the fiducial result, the total number count increases to 434534, and the constraint from $N$ has been improved by a factor of  $\sim$1.7 (1.1) in the self-calibration (known SR) scenario for both cosmological models.
However, the improvement for the constraint from $P$ is very limited, with a factor of only $\sim$1.02 in all cases, probably due to the large shot noises in the higher redshift bins.
Since the constraint on $M_\nu$ is dominated by $P$, the improvement in $\sigma^{N+P}(M_\nu)$ is not significant either. In the self-calibration (known SR) scenario, the improvement is only $\sim$1.1 (1.03) for both cosmological models.

We also investigate a wide range of $z_{max}$, results showing in \autoref{fig:zmax}.
The upper panel shows $\sigma(M_\nu)$ in $\nu \Lambda $CDM model. 
$\sigma^N(M_\nu)$ in the self-calibration scenario is too large to be shown in this figure, but it decreases significantly with the increase of $z_{max}$.
Conversely, $\sigma^P(M_\nu)$ remains relatively a constant after $z_{max}$ has reached $\sim$1.4. This constancy arises from the sharp drop in the cluster number density $\bar n$ Eq.\eqref{eq:barn} with increasing redshift, resulting in a very small effective volume $V_\text{eff}$ Eq.\eqref{eq:Veff} at high redshifts. 
Consequently, the constraint from the combination $N+P$, i.e. $\sigma^{N+P}(M_\nu)$, exhibits a much more moderate decrease with increasing $z_{\text{max}}$.

The lower panel shows the ratio of $\sigma(M_\nu)$ in $\nu w_0 w_a $CDM to $\sigma(M_\nu)$ in $\nu \Lambda $CDM model.
In the known SR scenario, the ratio from $N$ is significantly larger than 1 until $z_{max}$ reaches $\sim$2. This is because at low redshifts, dark energy can counterbalance the impact of neutrinos on the background evolution. However, at high redshifts, the influence of dark energy becomes secondary, refer to Fig.4 in \cite{Diaz_Rivero2019}.
On the other hand, the ratio from $P$ remains nearly constant once $z_{max}$ reaches $\sim$1.4, which is still due to the limit of $\bar n$.

\subsection{$k_{\text{max}}$}

\begin{figure}
	\centering	\includegraphics[width=0.9\linewidth]{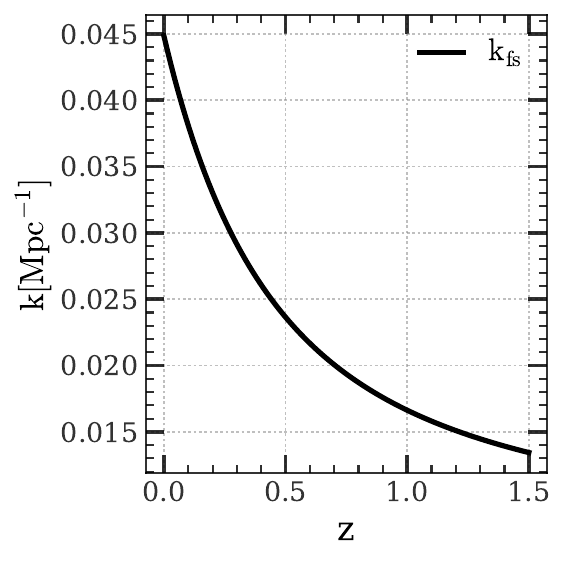}
	\caption{The free streaming scale \autoref{eq:kfs} as a function of redshift for $M_{\nu} = 0.06eV$.}
	\label{fig:kfs}
\end{figure}
\begin{figure}
	\centering	\includegraphics[width=0.9\linewidth]{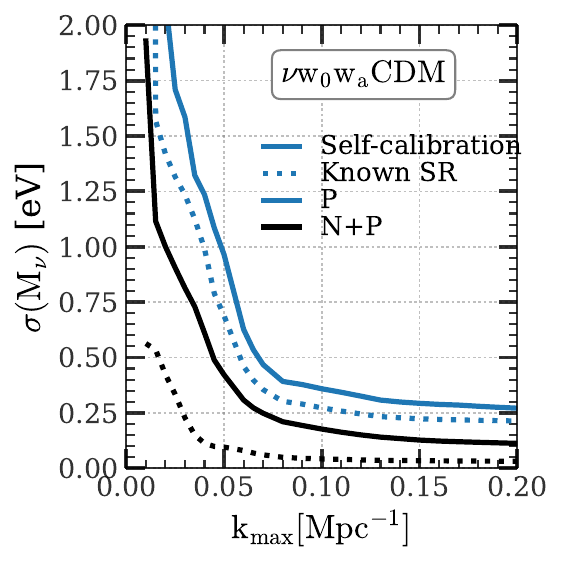}
	\caption{ 
    The impact of $k_{\text{max}}$ on $\sigma(M_\nu)$ for $\nu w_0 w_a$ CDM model.
    Different colors represent different probes: blue for cluster power spectrum $P$, and black for its combination with number counts $N+P$.
    Solid and dotted lines represent self-calibration and known SR scenarios respectively.}
	\label{fig:kmax}
\end{figure}

In our fiducial analysis for the cluster power spectrum, we have restricted to the linear regime by setting $k_{\text{max}} = 0.15$ \Mpcc \cite{Hu2003, Wang2005}.
Neutrino free streaming affects the matter power spectrum mainly on scales with $k\gsim k_{\text{fs}}$. \autoref{fig:kfs} shows $k_{\text{fs}}$ as a function of $z$ for the fiducial neutrino mass $M_{\nu}=0.06$eV. As can be seen, our range of $k$ modes covers the neutrino free-streaming scale over all redshift region, allowing us to probe the relative suppression of power on $k\gsim k_{\text{fs}}$ caused by massive neutirnos.
To assess the impact of specific choices of $k_{\text{max}}$, we gradually reduce $k_{\text{max}}$ from $0.15 $ \Mpcc down to $0.01 $ \Mpcc, with the corresponding constraints on $M_{\nu}$ shown in \autoref{fig:kmax} and the 7$^{th}$ row of \autoref{tab:resultDis}.

We find that cutting $k_{\text{max}}$ in half from $0.15$ \Mpcc to $0.07$ \Mpcc greatly weakens the constraint on $M_\nu$ from the power spectrum alone, with $\sigma^P(M_\nu)$ increasing by a factor of $\sim$1.6. The combined constraint from $N+P$ is also degraded, with $\sigma^{N+P}(M_\nu)$ increasing by a factor of about 1.6 $\sim$ 1.9.
When $k_{\text{max}}$ is set below $\sim0.04 $ \Mpcc, the average free streaming scale for the lowest redshift bin $[0,0.2]$, the constraint gets worse precipitously. This finding highlights the importance of choosing $k_{\text{max}}>k_{\text{fs}}$ and incorporating scales as small as possible when constraining $M_\nu$ with cluster power spectrum.

\subsection{Redshift uncertainty}\label{sec:DisSigmaz}

\begin{figure}
	\centering
	\includegraphics[width=0.9\linewidth]{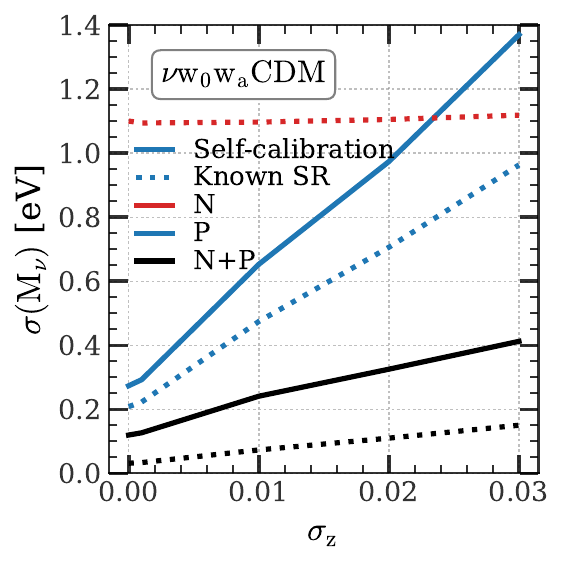}
	\caption{ The impact of redshift uncertainty parameter $\sigma_z$ on $\sigma(M_\nu)$ for $\nu w_0 w_a$ CDM model.
    Different colors represent different probes: red for the cluster number counts $N$, blue for cluster power spectrum $P$, and black for their combination $N+P$.
    Solid and dotted lines represent self-calibration and known SR scenarios respectively.
	}
	\label{fig:Sigmaz}
\end{figure}

The redshift uncertainty parameter $\sigma_z$ is optimistically assumed to be 0.001 in our fiducial analysis. This accuracy highly relies on the availability of accurate spectroscopic redshifts for all clusters observed by the CSST. In case it is challenging to achieve such an accuracy, we consider a more conservative estimate of 0.01 for $\sigma_z$. The constraint using this conservative estimate is presented in the 8$^{th}$ row of \autoref{tab:resultDis}. We also extend $\sigma_z$ up to 0.03 with results displayed in \autoref{fig:Sigmaz} for the $\nu w_0 w_a $CDM specifically.

Similarly to \autoref{fig:zmax}, $\sigma^N(M_\nu)$ in the self-calibration scenario is too large to be displayed in this figure. For both scenarios, the redshift uncertainty $\sigma_z$ has a minimal impact on $\sigma^N(M_\nu)$ due to the relatively large redshift bin size.
However, the influence of $\sigma_z$ on $\sigma^P(M_\nu)$ is noteworthy. This is attributed to the damping effect caused by the redshift uncertainty, which suppresses $P$ on relatively smaller scales. The behavior of $\sigma^{N+P}(M_\nu)$ with increasing $\sigma_z$ is similar to that of $\sigma^P(M_\nu)$, but less significant. Quantitatively, the ratio between $\sigma^{N+P}(M_\nu)$ with $\sigma_z=0.01$ and $\sigma_z=0.001$ is $\sim$2.
These findings underscore the importance of precise redshift measurements in effectively constraining the total neutrino mass.


\subsection{HMF uncertainty}\label{sec:DisHMF}

Although the CDM prescription we adopt in \autoref{sec:prescription} has  alleviated the impact of massive neutrinos on the universality of HMF, the HMF of Tinker et al (2008) we employ is accurate at the $\sim$5\% level \cite{Tinker2008}. Additionally, systematics such as different halo finders, non-$\Lambda$CDM simulations, baryonic effects etc. introduce extra uncertainty to the HMF \cite{Bhattacharya2011, Cui2012, Cui2014}. To account for these uncertainties, we incorporate two additional parameters $\{s, q\}$ to characterize the deviation of the Tinker's HMF from the true HMF, following \cite{Costanzi2019_preparation} and \cite{Abbott2020}:
\begin{equation}\label{eq:hmf_error}
	\frac{dn}{d \ln M} = \left( \frac{dn}{d \ln M} \right)_\text{Tinker} \left[s\log\left(\frac{M}{10^{13.8} \hMpc }\right)+q \right].
\end{equation}
The fiducial values are $s=0$ and $q=1$, reducing to the Tinker's mass function.

The 9$^{th}$ row in \autoref{tab:resultDis} shows the constraints with additional HMF uncertainties. 
Taking the $\nu w_0 w_a $CDM model as an example, compared to $\sigma(M_\nu)$ in the fiducial case, the constraints from $N$ have been relaxed by a factor of 1.2 (2.7) in the self-calibration (known SR) scenario, while the constraints from $P$ remain largely unchanged.
However for the constraints from $N+P$, the degradation becomes more significant by a factor of 1.85 (3.97).
For the $\nu \Lambda$CDM model, $\sigma(M_\nu)$ from $N+P$ degrades by a factor of 1.09(2.13).
As a result, the accuracy of the HMF plays a crucial role in constraining $M_\nu$ using $N+P$. It is worth mentioning that emulators for HMF, based on simulations with massive neutrinos, have the potential to achieve greater precision, bypassing the reliance on a universal formula \cite{Bocquet2020}.

\subsection{Bias systematics}\label{sec:DisBias}

Galaxy clusters are biased tracers of matter field. Fortunately, compared to galaxies, we have theory to describe the bias for clusters, and the formula can be calibrated with simulations, similar to the HMF. The bias of Tinker et al (2010) that we utilize already exhibits uncertainties.
Moreover, in real surveys, extra bias will be introduced by cluster selection effects. For example, the redMaPPer clusters show an additional mass-dependent bias due to projection effects and orientation biases \cite{To2021,Wu2022,Sunayama2020}.
We introduce two parameters $\{b_{s0}, b_{s1}\}$ to model systematics on cluster bias, following \cite{To2021,To2021PRL}:
\begin{equation}\label{eq:bias_error}
	b(M,z) = b_\text{Tinker}(M,z)b_{s0}\left(\frac{M}{5\times 10^{14} \hMsun}\right)^{b_{s1}},
\end{equation}
with fiducial values $b_{s0}=1,b_{s1}=0$.

The constraint on $M_\nu$ with the bias uncertainty is presented in the 10 $^{th}$ row of \autoref{tab:resultDis}.
For $\sigma^{P}(M_\nu)$, in the self-calibration scenario, the effect of bias uncertainty is completely negligible. 
However, in the known SR scenario, the inclusion of bias uncertainty relaxes the constraint by a factor of $\sim$1.03 (1.3) in the $\nu \Lambda$CDM ($\nu w_0 w_a$CDM) model.
On the other hand, for $\sigma^{N+P}(M_\nu)$, even in the self-calibration scenario, the effect of bias uncertainty is significant with a factor of $\sim$1.38(1.96). In the known SR scenario, it is 2.94 (3.5) in the $\nu \Lambda$CDM ($\nu w_0 w_a $CDM) model.
These findings highlight the importance of bias systematics to cluster cosmology, and emphasize the need for careful study of bias systematics in future analyses.

\subsection{GISDB}\label{sec:DisGISDB}
\begin{figure}
	\centering
	\includegraphics[width=0.9\linewidth]{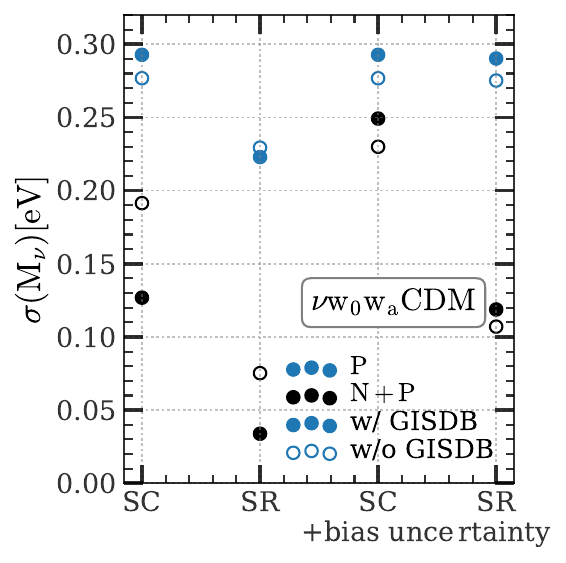}
	\caption{ 
    The impact of the GISDB effect on $\sigma(M_\nu)$ in $\nu w_0 w_a$ CDM model.
    Different colors (blue and black) represent different probes: cluster power spectrum $P$ and the combination $N+P$.
    Solid and hollow dots represent results with or without GISDB effect.
    From left to right, the columns represent the self-calibration scenario (SC), the known SR scenario (SR), SC with the bias uncertainty and SR with the bias uncertainty, respectively. 
	}
	\label{fig:Bias}
\end{figure}

In this subsection, we investigate the impact of the GISDB, which has been missed in previous cluster studies (e.g. \cite{Carbone2012, Sartoris2016, Cromer2019}). The constraint without GISDB is presented in the 11$^{th}$ row of \autoref{tab:resultDis}.

GISDB compensates partly for the suppression of the power spectrum caused by massive neutrinos on small scales. As a result, we expect that the constraint obtained from $P$ will be degraded when GISDB is considered. This is indeed what we find except for the ``known SR'' case in the $\nu w_0 w_a$CDM model, where $\sigma^{P}(M_\nu)$ gets slightly better from 0.229 to 0.223.
When using $P$ to constrain $M_\nu$, we utilize information not only from the shape of the power spectrum but also from its amplitude, predominantly through the effective bias $b_\text{eff}$.
By adding bias uncertainties, as discussed in the previous subsection, the amplitude information is blurred, leaving only the shape information.

Taking the $\nu w_0 w_a$CDM model as an example, according to \autoref{fig:Bias}, when bias uncertainties are added, $\sigma(M_\nu)$ with GISDB is consistently larger than \wen{that} without GISDB, regardless of whether the constraint is derived from $P$ or $N+P$. 
However, when the parameters $\{b_{s0}, b_{s1}\}$ are fixed, the results change. This is because the degeneracy between $M_\nu$ and $\{b_{s0}, b_{s1}\}$ differs when GISDB is considered or not.
Especially for $\sigma^{N+P}(M_\nu)$, considering GISDB will strengthen the constraint by a factor of $\sim$1.5 (2.2) in the self-calibration (known SR) scenario.

\subsection{Comparison with other forecasts}\label{sec:DisComp}

In this subsection, we compare our results forecasted for the CSST with other works in the literature forecasted for e.g. the LSST, Euclid, and CMB-S4.
All the constraints discussed here are derived for the $\nu w_0 w_a$CDM model, except for LSST, which is for the $\nu w_0$CDM model.

The first work using cluster number counts and cluster power spectrum to constrain the total neutrino mass was performed by Wang et al. (2005) for several cluster surveys including the LSST, which can provide a shear-selected cluster catalog in a sky area of 18,000 deg$^2$ and a redshift range of $z=[0.1,1.4]$. They forecast $\sigma^{P}(M_{\nu})=0.71$ eV and $\sigma^{N+P}(M_{\nu})=0.42$ eV, with an improvement factor of $\sim$1.70 when adding $N$.
In the self-calibration scenario, we obtain $\sigma^{P}(M_{\nu})=0.293$ eV and $\sigma^{N+P}(M_{\nu})=0.127$ eV, with a improvement factor of $\sim$2.31. 
This difference can be attributed to the fact that the total cluster number counts forecasted by Wang et al. is approximately $2\times 10^5$, which is much lower than our estimate of $4\times 10^5$.

Next, we compare our results with those obtained by Sartoris et al (2016) forecasted for the Euclid, which can provide an optical cluster catalog in a sky area of 15,000 deg$^2$ and a redshift range of $z=[0.2,2]$ \cite{Sartoris2016}. They use a selection threshold of $N_{500,c}/\sigma_\text{field}$, a ratio of $N_{500,c}$, the number of cluster galaxies contained within a certain radius, to $\sigma_\text{field}$, the rms of the field galaxy counts. They forecast a total cluster number counts of $2\times 10^6$ and $2\times 10^5$ for $N_{500,c}/\sigma_\text{field}=3$ and 5 respectively. Our cluster counts $4\times 10^5$ lies between these two numbers. 
For the former selection $N_{500,c}/\sigma_\text{field}=3$, they forecast  $\sigma^{N+P}(M_{\nu})=0.140$ eV in the self-calibration scenario. In the known SR scenario, it becomes 0.121eV with an improvement factor of $\sim$1.16.
Our main result is $\sigma^{N+P}(M_{\nu})=0.127$ eV in self-calibration, slightly smaller than $0.140$ eV. However, perfect knowledge of SR parameters brings our constraint greater improvement by $\sim$3.7 to $\sigma^{N+P}(M_{\nu})=0.034$ eV.
When extending $z_{max}$ from 1.5 to 2, our results become even smaller $\sigma^{N+P}(M_{\nu})=0.118\ (0.033)$ eV in self-calibration (known SR) scenario.
It is worth noting that if we do not consider the GISDB effect as they do, our result becomes $\sigma^{N+P}(M_{\nu})=0.191\ (0.075)$ eV, larger than their $0.140$ eV.

Finally, we compare our results with forecasts by Cromer et al (2019) for the CMB-S4, which can provide a Sunyaev-Zel'dovich (SZ) cluster catalog in a sky area of 10,000 deg$^2$ and a redshift range of $z=[0.1,1.9]$ \cite{Cromer2019}. They forecast a total cluster number counts of $\sim 10^5$. By combining with Planck primary CMB anisotropies and extra mass calibration from weak lensing, they obtain $\sigma^{N+P}(M_{\nu}) \sim 0.1 $ eV, which falls between our results of $0.127$ eV and $0.034$ eV in the self-calibration and known SR scenarios, respectively.
It is worth noting that they find the improvement on $\sigma(M_{\nu})$ when adding $P$ is very small $\sim$0.5\%. 
This may be due to the fact that they have already incorporated information from the primary CMB and additional mass calibration in cluster number counts, making adding in $P$ less effective. Furthermore, their redshift uncertainty is set to be a conservative value of 0.01, which has a significant impact on the constraining ability of $P$.
They also consider the bias uncertainty and find it has negligible effect on the combined constraint on $\sigma(M_{\nu})$, mainly because $P$ has minimal contribution to the constraint.

\section{conclusions}\label{sec:conclusions}

In this work, we forecast constraints on the total neutrino mass from the CSST galaxy clusters. Specifically, we employ the Fisher matrix formalism to derive the constraint from the cluster number counts $N$, cluster power spectrum $P$, and their combination $N+P$. 
In addition to considering the simplest extension to the $\Lambda$CDM model, the  $\nu\Lambda$CDM model ($\Lambda$CDM plus massive neutrinos), we also explore the inclusion of dynamical dark energy, the $\nu w_0 w_a$CDM model. We carefully take into account the effects of massive neutrinos on the halo mass function and halo bias.  
Since the largest source of systematics in galaxy cluster cosmology arises from the mass-observable relation, we examine two cases where the SR parameters are completely unknown or perfectly known, named as the self-calibration scenario and known SR scenario respectively. 
Our main results are presented in \autoref{tab:result} and \autoref{tab:resultDis}.
The results for the $\nu \Lambda CDM $ and $\nu w_0 w_a $CDM model are similar. Unless otherwise specified, the following results refer to the $\nu w_0 w_a $CDM model:

\begin{itemize} 

\item CSST is expected to detect $\sim400,000$ clusters with a mass threshold of $M_{200m} = 0.836 \times 10^{14} M_\odot/h$ in a survey area of 17,500 deg$^2$ and redshift range of $z=[0,1.5]$ , as depicted in \autoref{fig:abundance}.

\item $N$ provides a weak constraint on $M_\nu$, $\sigma^{N}(M_\nu) = 5.458\ (1.095)$ eV in the self-calibration (known SR) scenario. The constraint is dominated by $P$, with $\sigma^{P}(M_\nu) = 0.293\ (0.223)$ eV. However, when $N$ is added, the constraint can be improved by a factor of 2.31 (6.56), resulting in $\sigma^{N+P}(M_\nu)=$ 0.127 (0.034) eV.

\item Knowledge of the SR parameters has a significant impact on $\sigma^{N}(M_\nu)$ and consequently affects $\sigma^{N+P}(M_\nu)$. While partial knowledge leads to a small improvement in $\sigma^{N+P}(M_\nu)$ from $0.127$ eV to $0.125$ eV, having perfect knowledge results in a substantial improvement to $0.034$ eV. This corresponds to a $\sim 2\sigma$ detection of the minimum neutrino mass of $0.06$ eV.

\item Increasing $z_{max}$ from 1.5 to 2.0 will result in more detected clusters, increasing the total count from $4\times 10^5$ to $4.3\times 10^5$. While this change has a very small impact on $\sigma^{P}(M_\nu)$, improving it by a factor of 1.02 (1.02) in the self-calibration (known SR) scenario, it does strengthen $\sigma^{N}(M_\nu)$ by a factor of 1.64 (1.20). As a result, it strengthens $\sigma^{N+P}(M_\nu)$ a little bit, by a factor of 1.08 (1.03), leading to $\sigma^{N+P}(M_\nu) = 0.118 (0.033)$ eV.

\item We adopt $k_{\text{max}} = 0.15$ \Mpcc to avoid nonlinear effects. Reducing $k_{\text{max}}$ to a more conservative value of $0.07$ \Mpcc leads to a significant degradation in the constraint on $M_\nu$, with the uncertainty increasing by a factor of 1.6 for $\sigma^{P}( M_\nu)$ and 1.6 $\sim$ 1.9 for $\sigma^{N+P}(M_\nu)$.

\item Among the systematics we considered, the redshift uncertainty $\sigma_z$ has the largest impact on $\sigma^{P}(M_\nu)$, relaxing the constraint by a factor of 2.23 (2.13) when increasing $\sigma_z$ from 0.001 to 0.01. Consequently, this leads to a degradation in $\sigma^{N+P}(M_\nu)$ by a factor of 1.90 (2.15) in the self-calibration (known SR)  scenario.

\item We also examine the impact of uncertainties in the HMF and cluster bias. The HMF uncertainties primarily affect $\sigma^{N}(M_\nu)$, while the bias uncertainties only affect $\sigma^{P}(M_\nu)$. However, the effects of both uncertainties on $\sigma^{N+P}(M_\nu)$ are comparable, leading to a degradation of the constraints by a factor of approximately 2 (4) in the self-calibration (known SR) scenario

\item Finally, we investigate the impact of the GISDB, which has been missed in previous studies. This effect compensates for the suppression of the power spectrum caused by massive neutrinos on small scales, thereby one would expect it relaxes the constraint on neutrino mass. However, taking into account the degeneracy between $M_\nu$ and bias, as well as the addition of the probe $N$, our results show that considering this effect actually tightens $\sigma^{N+P}(M_\nu)$ by a factor of 1.50 (2.21) in the self-calibration (known SR) scenario.

\end{itemize} 

In summary, we find CSST clusters can provide a constraint on $M_\nu$ of $\sim0.1$ eV by combining N and P. With perfect knowledge of the SR parameters, this constraint can be tightened to $\sim0.03$ eV, enabling a $2\sigma$ detection of the minimum neutrino mass of 0.06 eV.
It is challenging to obtain perfect knowledge of the SR parameters. However, one can improve one's knowledge on the SR parameters by utilizing cluster weak lensing, for recent work on this topic, see e.g.\cite{eROSITA2024,SPT2024,LSST2025}. We plan to quantify the improvement by adding cluster lensing in future work. CSST also offers other probes to constrain neutrino mass.
For example, Lin et al. \cite{Lin2022} presented a forecast using mock data from the CSST photometric galaxy clustering and cosmic shear surveys, which yielded an upper limit for $M_\nu$ of approximately $0.23$ eV at a 68\% C.L. 
By combining different probes, it is expected that CSST can significantly strengthen the constraint on the total neutrino mass.
Moreover, the development of new probes such as voids \cite{Bonici2023} and Minkowski functionals \cite{Liu2022, Liu2023} has opened up new avenues for constraining neutrino masses. By combining these novel probes with traditional methods, it is expected that in the near future it will be possible to accurately constrain neutrino masses and distinguish between different mass hierarchies.

\begin{acknowledgments}
We thank Yan Gong for useful conversations. This work is supported by the National Key R\&D Program of China Grant No. 2022YFF0503404 and No. 2021YFC2203100, by the National Natural Science Foundation of China Grants No. 12173036 and 11773024, by the China Manned Space Program with Grant No. CMS-CSST-2025-A04, by the Fundamental Research Funds for Central Universities Grants No. WK3440000004 and WK3440000005, by Cyrus Chun Ying Tang Foundations, and by the 111 Project for "Observational and Theoretical Research on Dark Matter and Dark Energy" (B23042).
\end{acknowledgments}

\bibliographystyle{apsrev}
\bibliography{neuCSST}
	
\end{document}